\documentclass[fleqn,usenatbib]{mnras}

\usepackage{newtxtext,newtxmath}

\usepackage[T1]{fontenc}
\usepackage{ae,aecompl}
 
\usepackage{xcolor}
\usepackage{url}

\usepackage{graphicx}	
\usepackage{amsmath}	
\usepackage{siunitx}
\usepackage{ulem}

\usepackage{enumitem}
\setlist{leftmargin=7mm}

\title[GOTO Transient Recognizer]{Machine Learning for Transient Recognition in Difference Imaging With Minimum Sampling Effort }

\author[Mong et al.]
{Y. -L. Mong,$^{1,2}$\thanks{E-mail: yik.mong@monash.edu}
K. Ackley,$^{1,2}$
D. K. Galloway,$^{1,2}$
T. Killestein,$^{3}$
J. Lyman,$^{3}$
\newauthor
D. Steeghs,$^{3}$
V. Dhillon,$^{4}$
P. T. O'Brien,$^{5}$
G. Ramsay,$^{6}$
S. Poshyachinda,$^{7}$
\newauthor
R. Kotak,$^{8}$
L. Nuttall,$^{9}$
E. Pall\'e,$^{10}$
D. Pollacco,$^{3}$
E. Thrane,$^{1}$
\newauthor
M. J. Dyer,$^{4}$
K. Ulaczyk,$^{3}$
R. Cutter,$^{3}$
J. McCormac,$^{3}$
P. Chote,$^{3}$
\newauthor
A. J. Levan,$^{3}$
T. Marsh,$^{3}$
E. Stanway,$^{3}$
B. Gompertz,$^{3}$
K. Wiersema,$^{3}$
\newauthor
A. Chrimes,$^{3}$
A. Obradovic,$^{1}$
J. Mullaney,$^{4}$
E. Daw,$^{4}$
S. Littlefair,$^{4}$
\newauthor
J. Maund,$^{4}$
L. Makrygianni,$^{4}$
U. Burhanudin,$^{4}$
R. L. C. Starling,$^{5}$
R. A. J. Eyles-Ferris,$^{5}$
\newauthor
S. Tooke,$^{5}$
C. Duffy,$^{6}$
S. Aukkaravittayapun,$^{7}$
U. Sawangwit,$^{7}$
S. Awiphan,$^{7}$
\newauthor
D. Mkrtichian,$^{7}$
P. Irawati,$^{7}$
S. Mattila,$^{8}$
T. Heikkil\"{a},$^{8}$
R. Breton,$^{11}$
\newauthor
M. Kennedy,$^{11}$ 
D. Mata S\'{a}nchez,$^{11}$ 
E. Rol$^{1,2}$
\\ \\ 
$^{1}$School of Physics \& Astronomy, Monash University, Clayton VIC 3800, Australia\\
$^{2}$OzGrav: The ARC Centre of Excellence for Gravitational Wave Discovery, Clayton VIC 3800, Australia\\
$^{3}$Department of Physics, University of Warwick, Coventry, West Midlands, CV4 7AL, UK\\
$^{4}$Department of Physics and Astronomy, Hicks Building, The University of Sheffield, Sheffield, S3 7RH, UK\\
$^{5}$School of Physics and Astronomy, University of Leicester, University Road, Leicester, LE1 7RH, UK\\
$^{6}$Armagh Observatory \& Planetarium, College Hill, Armagh, BT61 9DB, Northern Ireland\\
$^{7}$National Astronomical Research Institute of Thailand,  260  Moo 4, T. Donkaew,  A. Maerim, Chiangmai, 50180, Thailand\\
$^{8}$Department of Physics and Astronomy, University of Turku, FI-20014 Turku, Finland\\
$^{9}$Institute of Cosmology and Gravitation, University of Portsmouth, Dennis Sciama Building, Burnaby Road, Portsmouth, PO1 3FX, UK \\
$^{10}$Instituto de Astrofisica de Canarias, La Laguna, Tenerife, Spain \\
$^{11}$Department of Physics and Astronomy, The University of Manchester, Oxford Road, Manchester, M13 9PL, UK
}

\date{Accepted XXX. Received YYY; in original form ZZZ}

\pubyear{2020}

\begin{document}
\newcommand{\Msun}{$M_{\odot}$}
\newcommand{\Lsun}{$L_{\odot}$}
\newcommand{\Rsun}{$R_{\odot}$}
\newcommand{\solar}{${\odot}$}

\label{firstpage}
\pagerange{\pageref{firstpage}--\pageref{lastpage}}
\maketitle

\begin{abstract}
The amount of observational data produced by time-domain astronomy is exponentially increasing. Human inspection alone is not an effective way to identify genuine transients from the data. An automatic real-bogus classifier is needed and machine learning techniques are commonly used to achieve this goal. Building a training set with a sufficiently large number of verified transients is challenging, due to the requirement of human verification. We present an approach for creating a training set by using all detections in the science images to be the sample of real detections and all detections in the difference images, which are generated by the process of difference imaging to detect transients, to be the samples of bogus detections. This strategy effectively minimizes the labour involved in the data labelling for supervised machine learning methods. We demonstrate the utility of the training set by using it to train several classifiers utilising as the feature representation the normalized pixel values in 21-by-21 pixel stamps centered at the detection position, observed with the Gravitational-wave Optical Transient Observer (GOTO) prototype. The real-bogus classifier trained with this strategy can provide up to $95\%$ prediction accuracy on the real detections at a false alarm rate of $1\%$.
\end{abstract}




\section{Introduction}
Transient astronomy focuses on astrophysical objects that vary on timescales of hours to years, and can originate from events such as supernovae, accreting binaries, stellar flares, tidal disruption events and gamma-ray bursts. Identifying and characterising transients is important for understanding astrophysics under extreme environments, accretion physics and the underlying physics of stellar flares.

In 2015, transient science stepped into a new era with the first direct detection of a gravitational wave (GW) event, GW150914 \citep{aaa16}, caused by the merger of a pair of $\approx30\ M_\odot$ black holes. Two years later, the first binary neutron star merger, GW170817, was detected \cite[]{gw170817}. GW detection alone can typically localize the event to only within a few hundred square degrees. To improve the localization down to order of an arcsecond, rapid-response electromagnetic follow-up observations are required \citep[e.g.][]{cfk17}. The identification of electromagnetic counterparts to the GW events is key to understanding the environments of the post-merger remnants \citep{met17}.

All-sky optical surveys can provide a more complete investigation of the optical transient sky. Time-domain astronomy has become a fast-growing area of astrophysics requiring comprehensive rapid-responsive strategies for following up the triggers of interesting events, such as gamma-ray bursts (GRBs) and GW events.

The recent advances of transient astronomy have been well-established by many transient survey projects, such as the SDSS-II Supernova Survey \citep{fbb08}; the Catalina Real Time Transient Survey \citep[CRTS,][]{ddm09}; Pan-STARRS1 \citep[PS1,][]{kbc10}; the Zwicky Transient Facility \citep[ZTF,][]{mas18}; the Asteroid Terrestrial-impact Last Alert System \citep[ATLAS,][]{tdh18}; and the SkyMapper Transient Survey \citep{wol18}, among others. In the future, of order $10^6$ transients are expected to be discovered per night with the \textit{Vera C.~Rubin Observatory} \citep{izk19}.

The Gravitational-wave Optical Transient Observer\footnote{\url{https://goto-observatory.org/}} (GOTO) is a robotic ground-based optical telescope located at the Roque de los Muchachos Observatory on La Palma, Canary Islands (Steeghs et al. in prep.). It is dedicated to searching for the optical counterparts to GW events. The GOTO prototype currently consists of $4\times40\,{\rm cm}$ unit telescopes (UTs) covering $\approx18\,{\rm deg}^2$ per exposure. The angular resolution of GOTO is about $1\farcs24$ per pixel. There are four Baader filters on each UT, a broad band $L$ filter (400--700\,nm), and narrower $B$, $G$ and $R$ filters. Under dark conditions the detection limit in the $L$ band is $\approx20.5\,{\rm mag}$ in 3 stacked $60\,{\rm sec}$ exposures. GOTO also performs an all-sky survey in order to discover other types of optical transients. The nightly sky coverage of GOTO is up to $\approx 2\,000\,{\rm deg}^2$.

To detect transients in an all-sky survey, difference imaging and ``real-bogus'' classification are the key steps. Difference imaging is the process under which a recently observed ``science'' image is subtracted from an earlier ``reference'' image for identifying excess flux  (see \S\ref{sec:imgprocess} for more details). However, as the difference images include both subtraction residuals and transient detections, real-bogus classification is required to separate them.

Due to a large number of detections (typically $\gtrsim10^4$) per GOTO image, source vetting and identification cannot rely solely on manual inspection. Efficient `real-bogus' classification on difference images has become one of the most important problems in transient astronomy, and several techniques have already been developed based on both supervised and unsupervised machine learning to address the problem.

There are two traditional ways to extract feature representations using supervised machine learning. Isophotal measurements of the detections (hereafter referred to as ``level-0'' attributes), could be used as the model features \citep{gbv17, brp13, brn12}.  Additionally, both linear and nonlinear combinations of level-0 attributes could generate more useful, but complicated features (hereafter referred to as ``level-1'' attributes). However, there are a vast number of ways to combine level-0 attributes, and trial and error tests have to be carried out in order to verify which level-1 attributes are useful. This ``feature engineering'' step becomes the most challenging part of the method. On the other hand, \citet{wss15} and \citet{gbv17}, hereafter referred to as W15 and G17 respectively, use pixel intensities as the feature representatives, which do not require any feature engineering.

Previous studies have shown that the learning algorithm and size of the training data set are the key factors affecting the performance of the classifier. W15 used a sample size of 32\,095 ($80\%$ training data, and $20\%$ test data).  \citet{brp13}, on the other hand, trained their classifiers on 50\,000 detections and tested the classifiers on a validation set with a size of 28\,448. The random forest (RF) technique is a machine learning algorithm with the architecture of multiple decision trees. It performed best in terms of the figure-of-merit (FOM) for both W15 and G17 studies i.e., using either isophotal measurements or normalized pixel values as the classification features. The FOM is defined as the minimum missed detection rate (MDR) with an acceptance of 1\% false positive rate (FPR). A convolutional neural networks (CNN) is another machine learning algorithm which is now widely used for image recognition in many different fields. Unlike RF, CNN only adopts pixel values to be the learning features. Some authors \citep[e.g.,][]{cfe16, gbv17, crf17} have claimed that CNN shows the best performance at picking out real candidates in difference images.

The most challenging part of applying supervised machine learning is in building up a sufficiently sized training data set in an automated way. Relying on human classification alone to create the training set is prohibitively expensive. Real transients in the data set could be defined as known transients identified by archival catalog searches or with prompt spectroscopy.
\citet{wss15} built up a dataset of $\approx8\,000$ real transients based on 3 years of Pan-STARRS1 observations, while \citet{brp13} used PTF observations taken in 2010 to build their training data set, where they identified 14\,781 real transients on difference images based on spectroscopy and other public domain data to create their real sample.

In this paper, we describe how we build a real-bogus classifier with minimum sampling effort. We begin with the motivation of this work followed by a brief description of the image processing in \S\ref{sec:motivation_imgprocess}. We describe the construction of our data sets and the feature extraction in \S\ref{sec:data_sets} and \S\ref{sec:feature_ex}. The models we use are described in \S\ref{sec:intro_classifiers}. In \S\ref{sec:results}, we compare the performance of our ``quick-build'' classifier (QB-classifier) with the one trained on an injection set (IT-classifier). Finally, we summarize our work in \S\ref{sec:conclusion}.

\section{Motivation and Image processing}\label{sec:motivation_imgprocess}
\subsection{Motivation}\label{sec:motivation}
The most straightforward approach to building a sample of real transients in the training set is to manually separate these from the few thousand bogus detections in each difference image (see \S\ref{sec:imgprocess} for more details on difference imaging). Using information from other transient surveys with spectroscopic classification, we can ensure that our sample of real transients is pure. However, there are two main problems with this approach. First, each of the samples in the data set has to be classified manually, which is a labour-intensive exercise. Secondly, it takes a long time to build a large data set, and the exercise is not easily scalable to even larger datasets.

To solve these problems, we have to understand how real detections appear on difference images. Unlike real detections, bogus artifacts typically do not appear as point-sources in the difference images. Consequently, one can reasonably assume that genuine transients in the difference images should have similar properties to the point-sources in the science images, since both detections can be described by a PSF superimposed on top of background noise. We can therefore create our training data set by collecting the training sample from the detections on the science images rather than from the difference images. This method of assembling a training data set does not require any human inspection allowing us to easily build up a very large sample.

There are several potential contaminants in the resulting sample: extended objects, such as galaxies, and artifacts, including cosmic rays, and hot pixels. The contaminant fraction can be reduced by filtering the outliers from the normalized full-width half-maximum (FWHM) distribution, and by using the \texttt{SExtractor} parameters CLASS\_STAR and ISOAREA\_IMAGE to exclude the galaxies and hot pixels from the real sample (see \S\ref{subsec:qb_train} for more details on \texttt{SExtractor}).

In parallel to the approach we used to build our real sample, we build our bogus sample by collecting all detections on the difference images.
Since we label all detections on the difference images as bogus, there may be some genuine transients included in the bogus sample. The contamination fraction in the bogus sample is estimated to be less than 1\% by assuming no more than 20 transients on each field.

With a large training data set, the machine learning model is less likely to be overfitted. Therefore, the decision boundary should be smooth enough to reject the outliers, which are the contaminants in our training set. As a result, we can maintain this negligible contamination in both real and bogus samples.

The key aim of this work is to demonstrate that our method of creating the training set is not only effective, but also easily applicable to different machine-learning algorithms to solve the real-bogus classification problem. We have therefore implemented different algorithms into the classifier to verify the feasibility of our approach.

\subsection{Image processing}\label{sec:imgprocess}
Raw images taken with GOTO are reduced automatically with our standard pipeline before performing further analysis (Ackley et al. in prep.). The standard pipeline applies bias correction, dark-frame subtraction and flat-field correction, followed by co-adding $3\times60\,{\rm sec}$ individual exposures to form a median science image. Throughout this study, we performed all analyses using \texttt{median} images as these have a higher signal-to-noise ratio than individual exposures.
\indent The {\tt template} image, also referred to as the reference image, is a previous image of the same field that is subtracted from all successive science frames. Since GOTO operates by tiling the sky on a fixed grid \citep{ddl18}, we are able to update the set of templates regularly.

Image alignment and difference imaging are part of the standard pipeline procedures following calibration.
We use a modified version of the Python package \texttt{alipy} to align the template image to the science image by cross-matching positions of selected field stars using high-order affine transformations independent of the WCS information. Once the alignment has been performed, we use \texttt{HOTPANTS}\footnote{\url{https://github.com/acbecker/hotpants}} \citep{bec15} to perform image subtraction.

\section{Data sets}\label{sec:data_sets}
We use three data sets  in this work: the quick-build training set, the injection data set for both testing and training, and the minor planet (MP) test set (Table \ref{tab:data_set}).

The quick-build training set is used to train our real-bogus classifier. We apply our quick-build strategy which can effectively assemble real detections in our training set. In practice, we are primarily concerned about the performance of the classifiers applied on the difference images. Since all the real samples in this training set are collected from the science images, this data set will not be used for any testing purpose. Therefore, as we need a reliable test set for testing the performance of our classifiers, we are motivated to build the injection data set and MP test set.

The injection data set is generated by collating all of the detections from the difference images after performing difference imaging on the injected science images. We apply this data set in two ways. The first is to use the injection set to test the classifiers trained on the quick-build training set. On the other hand, since the morphology of the injections are a good representative of how genuine transients may appear in practice, we also use the injections recovered on the difference images to train our classifier. This, in effect, mimics the training process using genuine transients in the standard way and will indirectly provide a figure of merit comparison with the classifiers trained using the quick-build method.

We consider the MP test set to be the most reliable test set, over the injection test set, as we use verifiable MPs as our real sources. The classifiers trained on the quick-build and the injection data sets will be tested on this MP test set for performance evaluation and to provide evidence of the efficacy of our method.

\subsection{``Quick-build'' training set}\label{subsec:qb_train}
To ensure that the quality of the images used to build our data set is sufficiently high enough, we select images based on several criteria. We randomly select 45 science images between April to May 2019 from different fields taken with different UTs for building our real sample. We avoid choosing images where the number of detections are <15\,000 within the FoV ($=2.1\times2.8\,{\rm deg}^2$) of a single UT to ensure a large enough representation of samples.

\texttt{SExtractor} is commonly used to identify detections, which have a higher pixel counts as compared to the background level, in an image \citep{ba96}. The default sensitivity parameter of the {\tt SExtractor}, \mbox{DETECT\_THRESH}, is set to $2.0\sigma$ for science images and $2.5\sigma$ for difference images on GOTO standard pipeline.

In order to avoid bad detections in our real sample, such as those on the edge of the frame,  saturated or spurious pixels, etc., we filter out the detections with non-zero FLAGS\footnote{\url{https://sextractor.readthedocs.io/en/latest/Flagging.html}}. This step will remove saturated bright objects (FLAG=4), and any objects that are too close to bright objects (FLAG=2). For those objects next to bright objects that are well-deblended (FLAG=0), they are also included in our training set since they should still resemble a PSF on top of the background. We identify that flagged detections contribute $\approx10$\% of the entire real sample. We further reduce the contaminants by filtering the detections falling outside the range between 0.3\% and 99.7\% percentiles of the normalized FWHM distribution over each image, as well as detections brighter than $m$=12 were also removed in order to reduce the contamination due to bright objects with diffraction spikes. Finally, we build our real sample of 455\,673 objects purely using the detections extracted from the science images.

Similarly, we use 680\,775 detections extracted from 49 difference images  to build our bogus sample. There is a small fraction of true negative contaminants in the bogus sample due to the presence of real transients in the difference image. In most cases, $\approx10^{3-4}$ detections are recovered by \texttt{SExtractor} in a single difference image. Among them, there are typically fewer than 20 real transients per image, i.e. typically $<1\%$. For those frames aligning on the galactic plane, there could be a higher number of recovered variable sources. However, the bogus artifacts that arise due to image subtractions residuals or template mis-alignments greatly outnumber the number of variable sources or true transients, and generally scale with the density of sources in the field. Therefore, the contamination fraction still remains less than 1\%. Building our bogus sample using all of the detections on the difference image (less than 1\% contamination) without human inspection is acceptable if the sample size is large enough. Combining with the real sample, our entire data set contains 1\,136\,448 detections.

To ensure our training set is balanced, we randomly select 400\,000 detections from each of the real and bogus samples, to form our training set for a total size of 800\,000 detections.

\subsection{Injection data set}\label{subsec:inj_set}
We create another data set by using images with simulated sources injected into them. We use this data set both for testing the performance of the classifiers trained on the quick-build training set (see \ref{subsec:qb_train}), and to train another classifier for comparison purposes.

We use the field of SN2019pjv located at $\alpha$=17:14:34.8, $\delta$=+28:07:26.1 (J2000), which has been revisited by GOTO 91 times on different nights between September 2019 and February 2020, as our injection field. Since UT3 and UT4 were relatively stable, in terms of the FWHM compared to other UTs, we select images with QUALITY\_FLAG=0, for which the quality assessment of the images is calculated (see Ackley et al. in prep.), to perform injections, resulting in a total of 143 injected images.

We perform the injection process using \texttt{IRAF} \citep{tod86, tod93}. We uniformly inject point sources over each image, with apparent magnitudes in the range $m$=15--21. The total number of injections which are recovered by {\tt SExtractor} after difference imaging is 70\,891, giving a 63\% recovery rate.

We define all 70\,891 injections on the difference images to be the real sample of our training set. Furthermore, we double our real sample by reflecting all injection stamps along the diagonal image axis in order to create a larger data set. To build a balanced data set, we sample 141\,782 bogus detections randomly for our bogus sample. In sum, our entire injection training set contains 283\,564 detections.
\begin{table}
	\centering
	\caption{Number of detections in different data sets}
	\label{tab:data_set}
	\begin{tabular}{lccc}
		\hline
        \textbf{Data sets} & \textbf{Bogus} & \textbf{Real} & \textbf{Total}\\
        \hline
        Quick-build training set & 400\,000 & 400\,000 & 800\,000 \\
        Injection data set & 141\,782 & 141\,782 & 283\,564\\
        MP test set & 42\,929 & 33\,511 & 76\,440 \\
		\hline
	\end{tabular}
\end{table}
\subsection{Minor Planet (MP) test set}\label{subsec:test_set}
As a representative example of on-sky performance for genuine transient sources, we assemble a test set using archival MP detections from the past year of GOTO operations. This data set has the benefit over an injection set for accurately sampling across a wide range of field densities, image PSFs, and sky conditions. MP detections have similar properties to those of genuine transient objects --- they are detected in the science image, but absent in the template image, due to the large sky motion of the object, which leaves a ``clean'' subtraction residual and is similar to what we expect from genuine transient sources.

To build this test set, we randomly select 12\,000 images from the GOTO database. For each image we obtain the positions of all known MPs within the field of view using the SkyBoT cone search \citep{berthier06}. These positions are then cross-matched with the difference photometry table of each image, to identify the detected MPs in each image. We adopt a cross-match radius of $1\farcs$, to minimise contamination from spurious associations. To generate a matching bogus sample for the test set, we randomly sample from the difference image detections, choosing one for every MP detected per image. This approach provides an unbiased sample of the typical bogus content of each image, and due to the significant imbalance between real and bogus detections, provides a largely clean bogus sample.

The largest source of contamination within this sample is variable stars. Inevitably, when selecting a random sample of sources in the image a small fraction of these will be variable, and could show a clear residual in the difference image, depending on the amplitude of variability. Those with clear residuals will have incorrect (bogus) labels and be marked as misidentifications in the training set due to the classifier scoring them as real. These contaminants would negatively skew any performance metrics calculated. Determining algorithmically which labels to assign these detections is difficult, and is likely to inject bias. We opt to remove all variable stars from the training set. After generating the test set, we cross-match the coordinates of the random bogus sample against the ATLAS Variable Star Catalogue \citep{heinze18}, with a generous cross-match radius of $5\farcs$. This aims to maximise completeness in removal of variable stars, at the cost of some non-variable objects being removed. Typically around 4\% of the test set is removed with this cut. 

As a final cut, we remove cosmic rays from this test set. These features cannot always be distinguished in the difference image alone, because when \texttt{HOTPANTS} convolves the science image with the PSF kernel, these detections become PSF-like. We reject detections that only have one detection in the individual images that form the median. We opt for this approach to avoid removing sources that are undetected in the individual images due to poor signal to noise ratio, yet appear in the median stack. 

Applying all of the steps above results in a test set of $\approx76\,440$ examples, with the ratio $1:1.6$ MPs to random bogus detections. Our methodology for automated test set production is detailed more thoroughly in Killestein et al. (in prep.).

\section{Feature extraction and preprocessing}\label{sec:feature_ex}
\begin{figure}
	\includegraphics[width=\columnwidth]{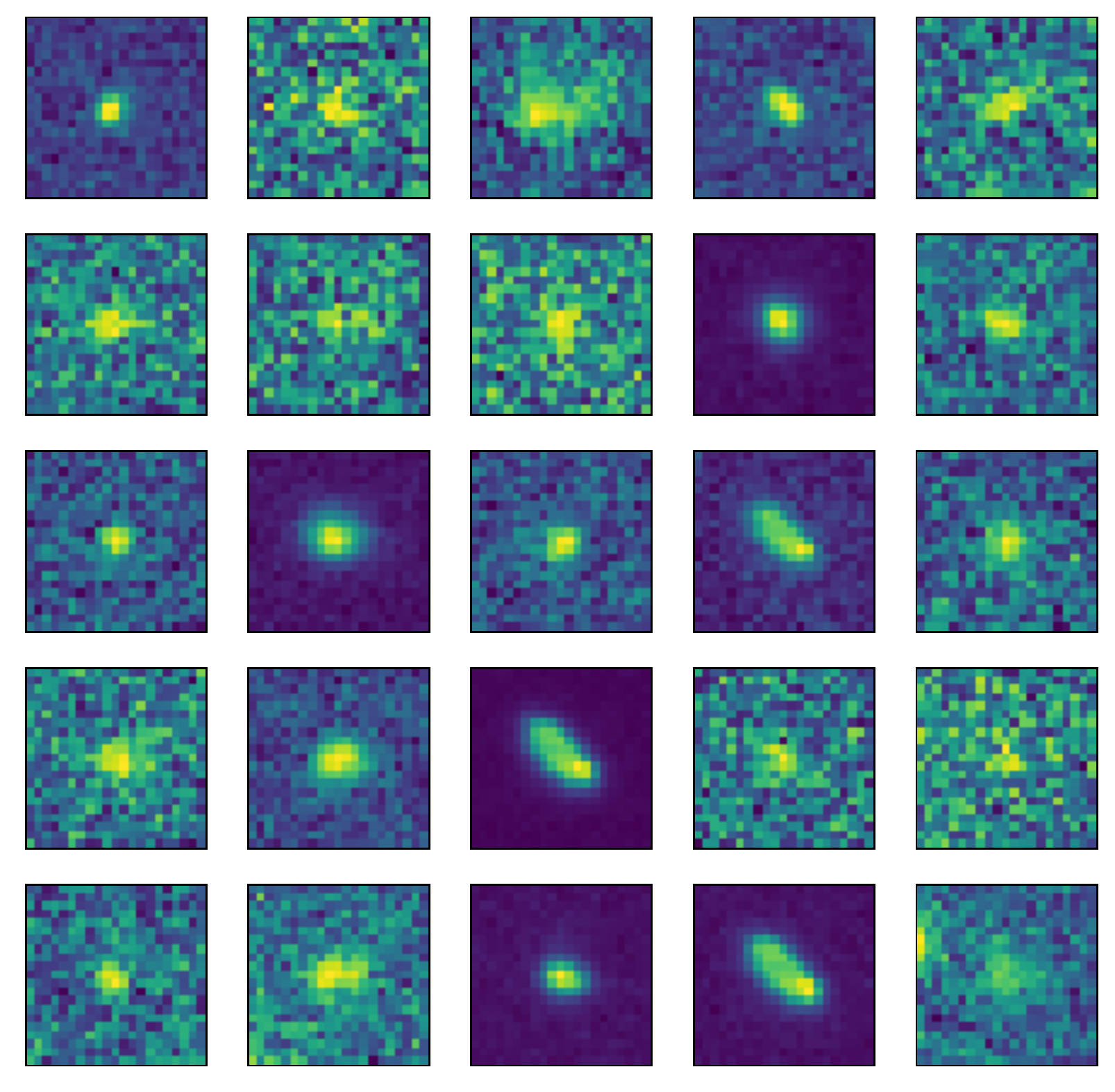}
	
	\vspace{.5cm}
	
	\includegraphics[width=\columnwidth]{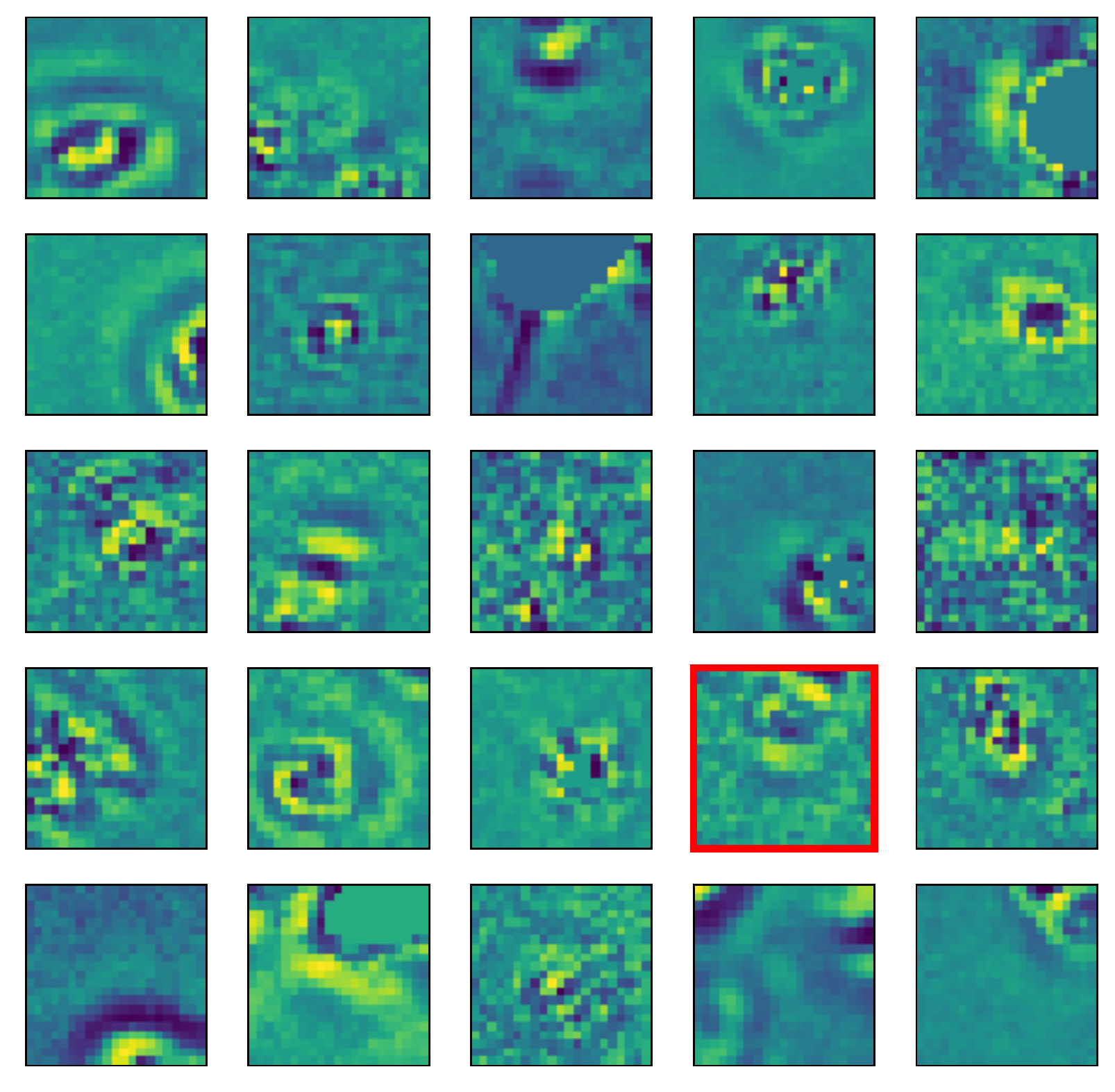}
    \caption{Examples of the 21-by-21 pixel thumbnails in the training set. \textit{Top five rows}: examples of  real detections. \textit{Bottom five rows}: examples of  bogus detections. The \textit{red} framed bogus stamp shows the segmentation of detection in the image subtraction process.
    }
    \label{fig:example_stamps}
\end{figure}

To extract the pixel intensity features, we crop a 21-by-21 pixel (26-by-26 arcsec) stamp centred at the image coordinate (\texttt{X\_IMAGE}, \texttt{Y\_IMAGE}) of the detection as measured by \texttt{SExtractor} for each sample in the training set (see Fig.~\ref{fig:example_stamps} for some examples). The real detections are all located at the center of the stamp with a typical aperture size of $\approx 5\farcs$ surrounded by shot noise. On the other hand, the segmentation of the subtraction residual might occur such that {\tt SExtractor} would identify multiple bogus detections for a single astronomical object. The \textit{red} framed bogus stamp in Fig. \ref{fig:example_stamps} is an example showing that a single object is segmented into three detections after difference imaging. It typically results in an offset between the segment of each subtraction residual and the actual position of the source.

Due to the appearance of masked pixels and missing values (off-edge pixels) within the pixel stamp in some cases, data cleaning was necessary before performing further analysis. During the subtraction of bright sources, masked pixels can be generated in the difference image (e.g. see the bottom left thumbnail in Fig.~\ref{fig:example_stamps}).
We clean the data by replacing all masked and off-edge pixels by the median value of the stamp, which is approximately the background level. 

Since each detection has its own signal-to-noise level relative to the background noise, we normalize the pixel intensities with
\begin{align}
    f(p_i) = \frac{p_i-\tilde{p}}{|p_i-\tilde{p}|}\log_{10}\left(1+\frac{{|p_i-\tilde{p}|}}{\sigma}\right)~,
\end{align}
where $p_i$ is the $i$-th pixel value. $\tilde{p}$ and $\sigma$ are the median and the standard deviation of the pixel intensities in the stamp. This scaling algorithm is adopted from W15 and \texttt{EYE}\footnote{\url{http://www.astromatic.net/software/eye}} \citep{ber01}, with the modification that $p_i$ is replaced by $p_i-\tilde{p}$. In  previous studies \citep[e.g.,][]{gbv17,wss15,brp13,brn12}, the real sample was collected from the difference image, implying that the background level should always be around zero. In our case, since we use unsubtracted science image detection to comprise the real sample, the background level is always non-zero. Therefore, we reset the noise level at the median pixel value of the stamp.
\begin{table}
	\centering
	\caption{Model parameters of the ANN and the RF we adopt.}
	\label{tab:models}
	\begin{tabular}{lr}
		\hline
        \textbf{Model parameters} & \textbf{Values}\\
        \hline
        \multicolumn{2}{l}{\textbf{Artificial Neural Network}} \\
		Size of $1^{\rm st}$ \texttt{Dense} layer & 100 \\
		Activation (hidden layer) & \texttt{ReLu}\\
		Regularization & $ \lambda=0.03 $\\
		Optimizer & \texttt{RMSprop} \\
        \multicolumn{2}{l}{\textbf{Random Forest Regressor}} \\
        \texttt{n\_estimators} & 1000 \\
        \texttt{max\_features} & 25 \\
        \texttt{min\_samples\_leaf} & 1 \\
		\hline
	\end{tabular}
\end{table}
\section{Classification Algorithms}\label{sec:intro_classifiers}
We build our classifiers using two different supervised machine learning algorithms: the random forest \citep[RF,][]{randomforest} and the artificial neural network \citep[ANN,][]{ann}. These algorithms are selected due to their reasonable performance shown in the literature \citep{wss15}. We use the Python packages \texttt{sklearn} \citep{sklearn}, \texttt{keras} and \texttt{tensorflow} \citep{tensorflow} to build the RF and the neural network models, respectively.

We tune the hyperparameters of each model to optimize performance, and list the optimal hyperparameters in Table\,\ref{tab:models}. We build our single-layer ANN model with 100 neurons. Activation functions \texttt{ReLu} and \texttt{softmax} are used in the hidden layer and the output layer respectively. The optimizer we use in ANN is \texttt{RMSprop}. For our RF classifier, we build it with \texttt{n\_estimators}=1000, \texttt{max\_features}=25 and \texttt{min\_samples\_leaf}=1.

\section{Results and Performance}\label{sec:results}
In this section, we show the general performance of the classifiers trained on the different data sets.

In order to mimic a more realistic case of applying our classifier to difference images directly, we verify the efficacy of different learning algorithms by testing on the injection data set (see \S\ref{sec:inj_test}). We also compare the performance between the classifiers trained on the quick-build training set and the injection data set. In \S\ref{sec:mp_test}, we compare the performance of the classifiers trained on different data sets by testing them on our MP test set.
\begin{figure}
	\includegraphics[width=\columnwidth]{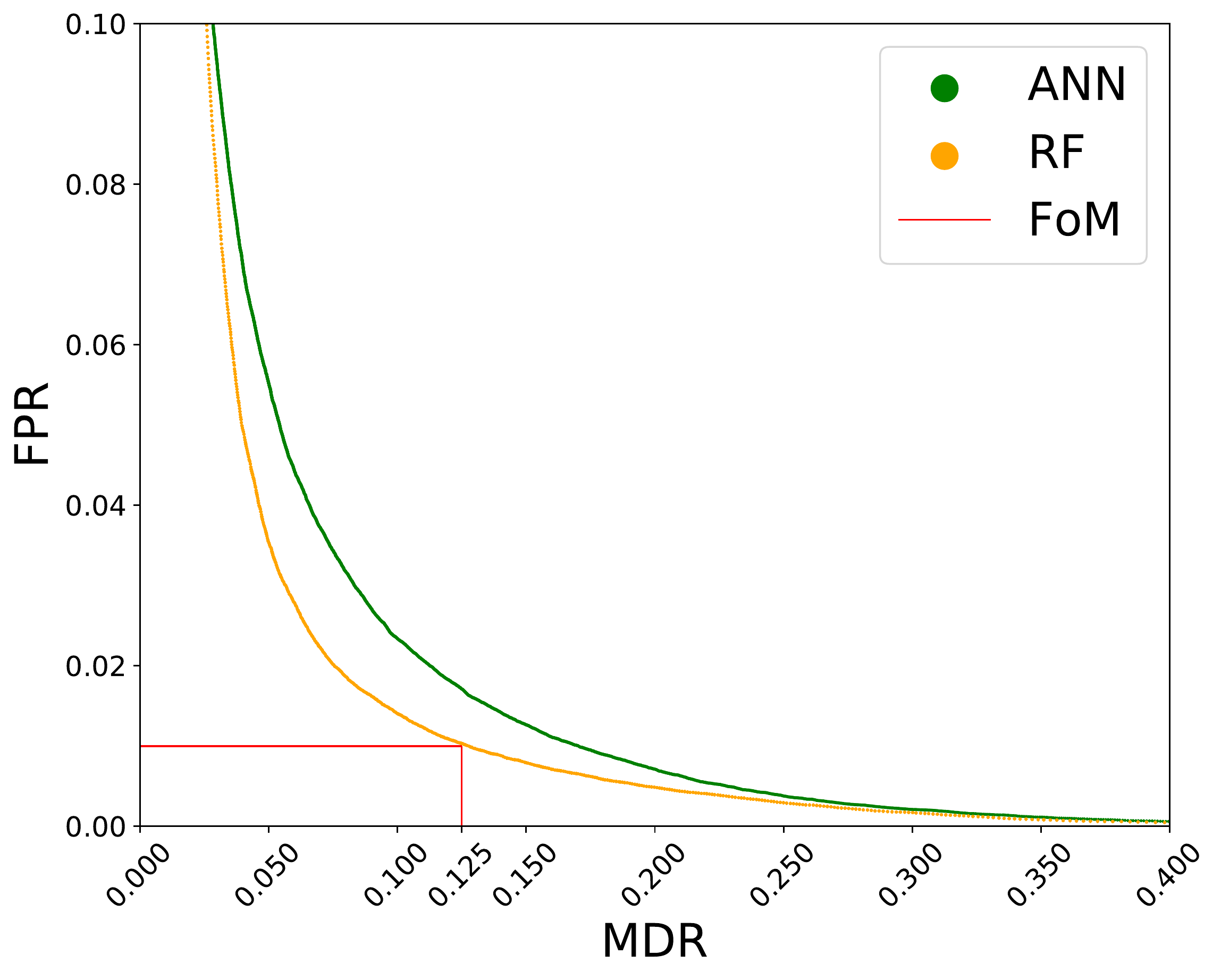}
    \caption{The receiver operator characteristic (ROC) curves of the injection test applied to different learning algorithms. The ANN and RF classifiers are represented by \textit{green} and \textit{orange} lines, respectively. The RF classifier shows a better performance, with  figure of merit (FoM,  indicated by the {\it red} line}) of 12.5~percent.
    \label{fig:roc}
\end{figure}
\begin{figure}
	\includegraphics[width=\columnwidth]{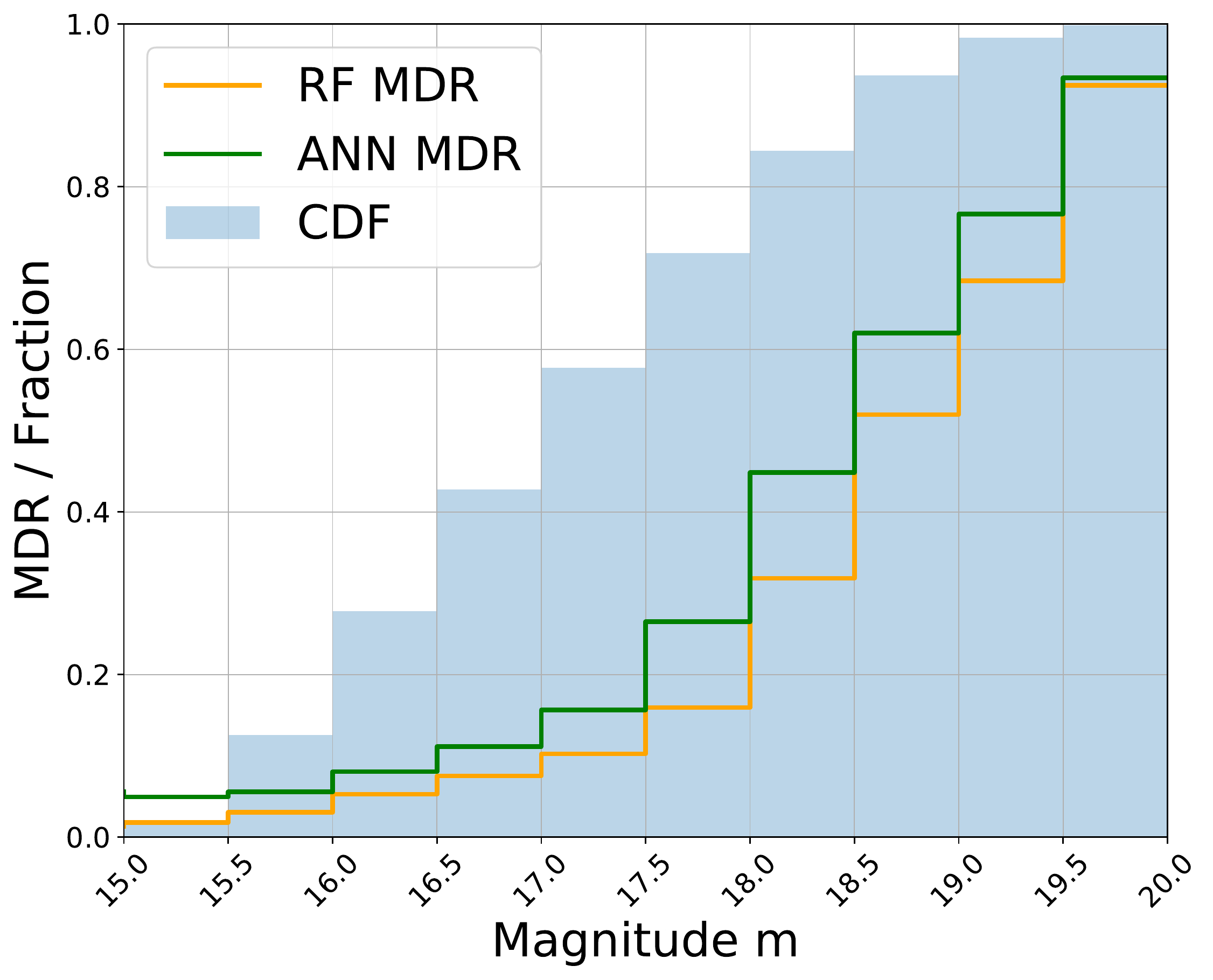}
    \caption{MDR for the injections as a function of magnitude. The RF classifier indicated by the \textit{orange} line always shows a lower MDR over the ANN model (\textit{green} line). The \textit{blue} blocks show the cumulative distribution function (CDF) of the magnitude of the injections.}
    \label{fig:mdr_mag}
\end{figure}

\subsection{Performance of the injection test}\label{sec:inj_test}
The injection data set consists of 283\,564 samples with a 1:1 balance ratio between the numbers of real and bogus detections. We label all injections as real detections and leave the rest as bogus. Therefore, since there are some real transients existing on the difference images which are not injections but are labelled as bogus, the false positive rate calculated from the injection test could be overestimated. With known magnitudes of all injections, we can study how the recovery rate would be affected by the brightness of the detection.

We compare the performance of ANN and RF models by plotting the receiver operator characteristic (ROC) curves (see Fig. \ref{fig:roc}). We conclude that the RF classifier performs better in terms of both area under the curve (AUC) and figure of merit (FoM).

We investigate how the MDR varies with the brightness of the detections in Fig. \ref{fig:mdr_mag}. The decision boundary is set to FPR=0.01 for each of the learning algorithms (see Table \ref{tab:compare_tab_inj}). The RF classifier has the lowest MDR over the range of magnitudes from m=15 to m=20. We also plot the cumulative density function (CDF) against the magnitude in Fig. \ref{fig:mdr_mag}. The constant step size of about 10 to 15\% from m=16 to m=18.5 in the CDF shows a uniform magnitude distribution of the injections in our data set. The decrease in the step size beyond m=18.5 is due to the drop of the \texttt{SExtractor} recovery rate with the increase of magnitude as we are nearing the limiting magnitude of GOTO.
\begin{table}
	\centering
	\caption{Decision boundaries and prediction accuracies at FPR=0.01 in the injection test}
	\label{tab:compare_tab_inj}
	\begin{tabular}{lcccc}
		\hline
        \textbf{Algorithms} & \textbf{Decision} & \textbf{Real} & \textbf{Bogus} & \textbf{FoM} \\
        & \textbf{boundary} & \textbf{accuracy} & \textbf{accuracy} & \\
        & & \textbf{$\%$} & \textbf{$\%$} & \textbf{$\%$} \\
        \hline
        RF & 0.75 & 87.5 & 99.0 & 12.5\\
        ANN & 0.91 & 83.9 & 99.0 & 16.1\\
		\hline
	\end{tabular}
\end{table}
\begin{figure}
	\includegraphics[width=\columnwidth]{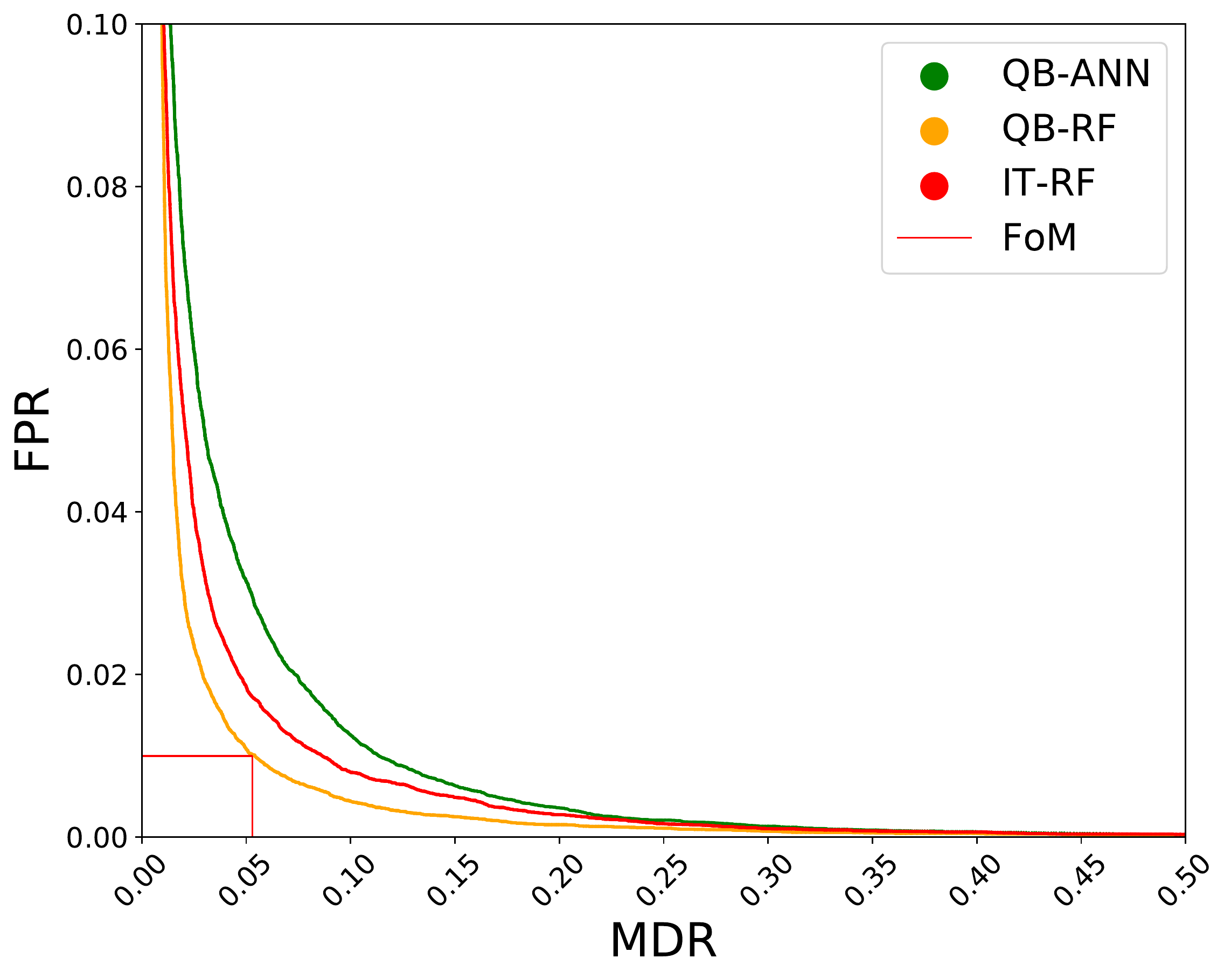}
    \caption{The ROC curves of different learning algorithms tested on the MP test set. QB-ANN and QB-RF classifier are represented by \textit{green} and \textit{orange} lines, respectively. The QB-RF classifier shows the best performance with the FoM of 5.2\%. The IT-RF classifier represented by the \textit{red} line shows a consistent performance with the QB-RF classifier.}
    \label{fig:mp_roc}
\end{figure}
\begin{figure}
	\includegraphics[width=\columnwidth]{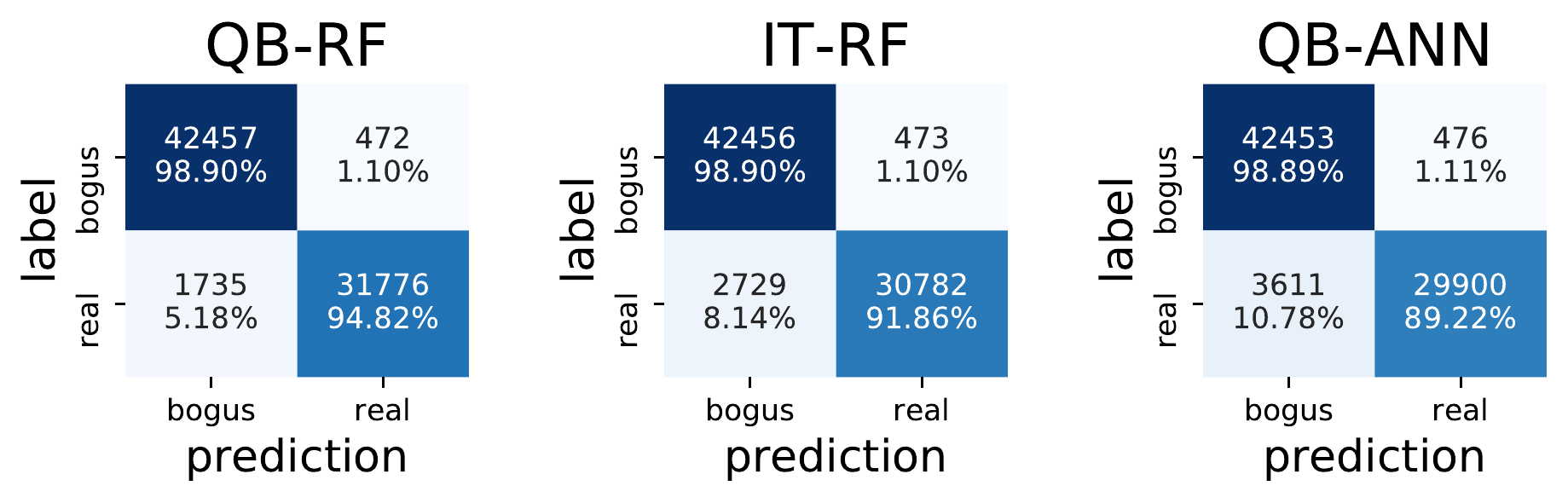}
    \caption{Confusion matrices of different models performing on the MP test set. The decision boundary of each classifier is set at FPR=0.01. The QB-RF shows the highest real prediction accuracy of $94.8\%$.}
    \label{fig:cm}
\end{figure}
\begin{figure}
    \centering
	\includegraphics[scale=0.5]{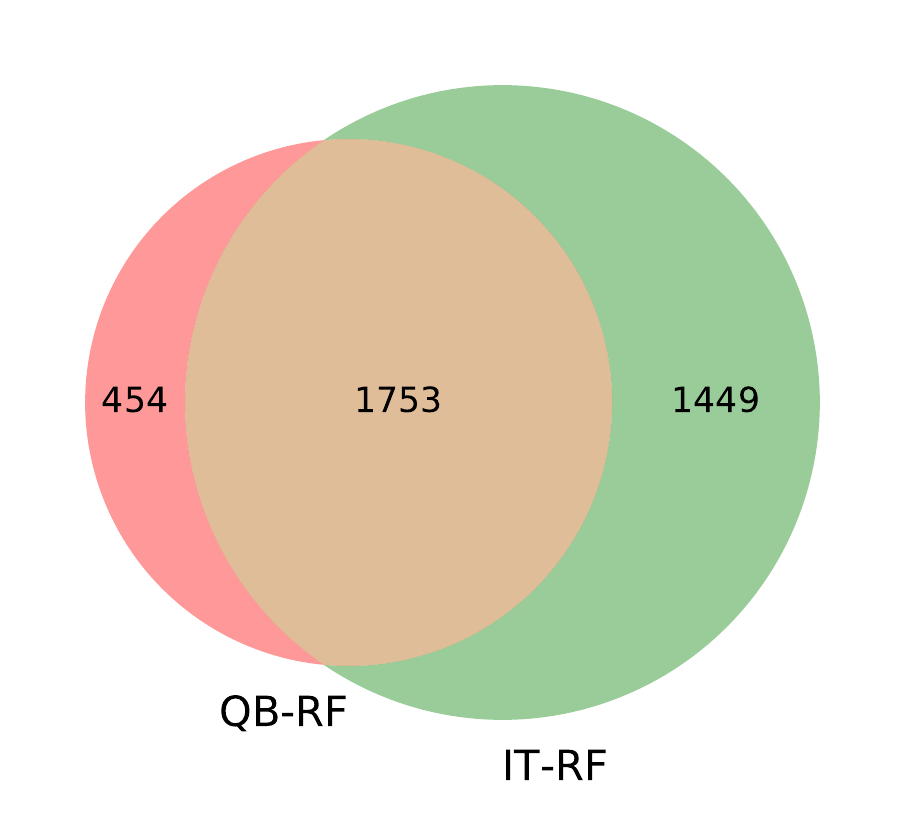}
    \caption{Venn diagram of the number of misclassified sources for QB-RF and IT-RF. It shows that $\approx80\%$ of the QB-RF misclassifications are also misclassified by IT-RF.}
    \label{fig:venn_diag}
\end{figure}
\begin{table}
	\centering
	\caption{Decision boundaries and prediction accuracies at FPR=0.01 testing on the MP test set}
	\label{tab:compare_tab}
	\begin{tabular}{lcccccc}
		\hline
        \textbf{Algorithms} & \textbf{Decision} & \textbf{Real} & \textbf{Bogus} & \textbf{FoM} &
        \textbf{F1}\\
        & \textbf{boundary} & \textbf{accuracy} & \textbf{accuracy} & & 
        \textbf{score}\\
        & & \textbf{$\%$} & \textbf{$\%$} & \textbf{$\%$} & \\
        \hline
        QB-RF & 0.61 & 94.8 & 99.0 & 5.2 & 0.97\\
        IT-RF & 0.55 & 91.9 & 99.0 & 8.1 & 0.95\\
        QB-ANN & 0.86 & 89.2 & 99.0 & 10.8 & 0.94\\
		\hline
	\end{tabular}
\end{table}

\subsection{Performance on the MP data set}\label{sec:mp_test}
In this section, we include one more RF classifier trained on the injection set (IT-RF) in our analyses. The purpose of comparing with the IT-RF classifier is to show that the classifiers trained on our quick-build training set also perform a consistently with the classifier trained on the data solely collected from the difference images.

We test our classifiers on real data by using our MP test set (see \S\ref{subsec:test_set}). According to the ROC curves in Fig. \ref{fig:mp_roc}, the RF classifier trained on the quick-build training set (QB-RF) shows the lowest FoM of 5.2\%. Both AUC and FoM also show that QB-RF and IT-RF perform consistently with each other. The decision boundaries used in this section and the FoMs of all classifiers are also showed in Table \ref{tab:compare_tab}. Since our MP test set is slightly unbalanced with real-to-bogus ratio of $1:1.3$, we also list F1 scores, which helps to estimate the goodness of balance between the recall and the precision, in Table \ref{tab:compare_tab}. The F1 score of QB-RF is closest to 1, indicating that this model is superior to the other models considered. We also show the confusion matrices for different classifiers at a fixed FPR of 1\%in Fig. \ref{fig:cm}. We plot the Venn diagram in Figure \ref{fig:venn_diag} to compare the misclassification consistency between QB-RF and IT-RF. The intersection is about $80\%$ of the QB-RF population, which implies the misclassifications of the two models are consistent with each other.

\begin{figure}
	\includegraphics[width=\columnwidth]{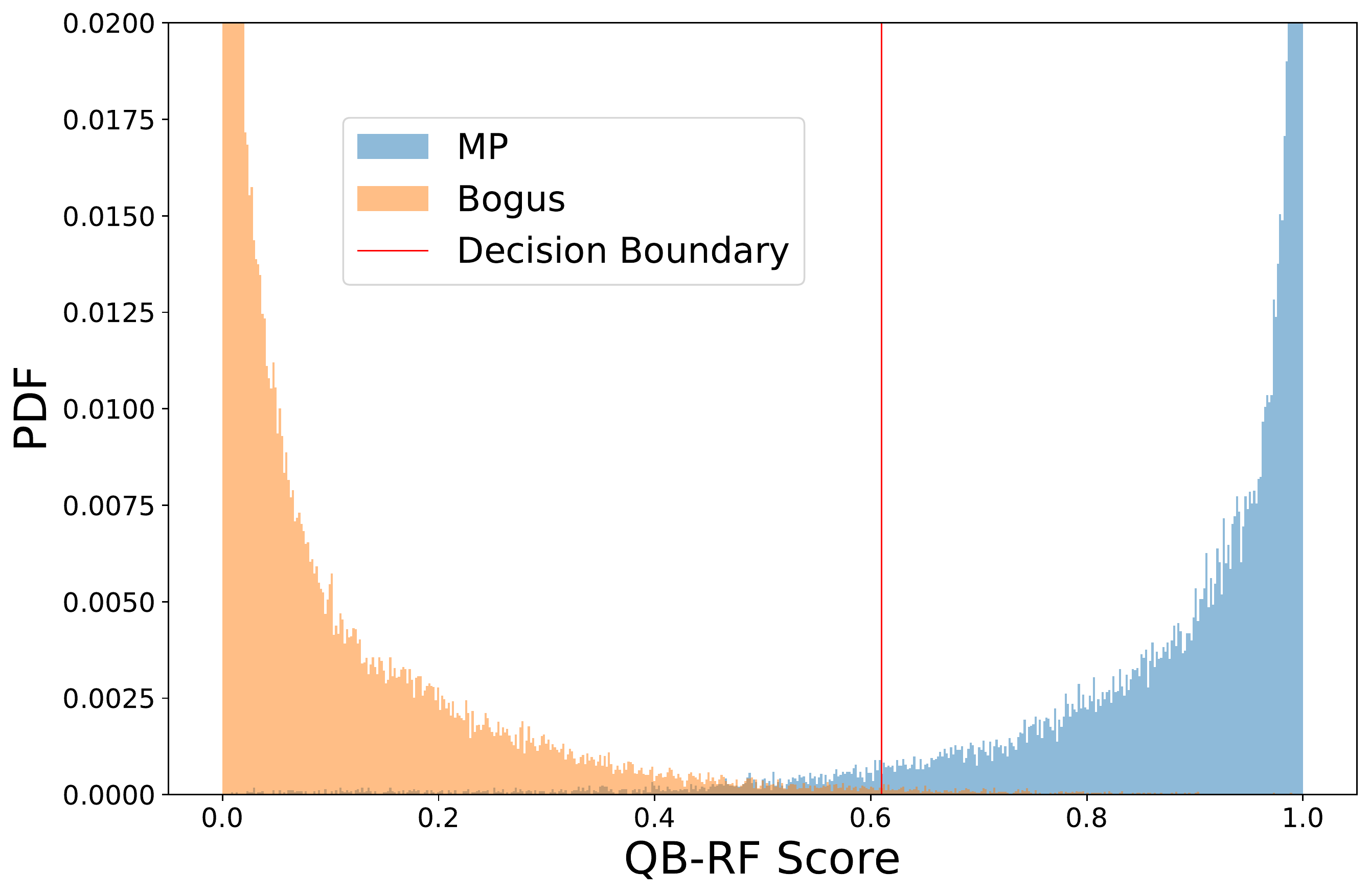}
	\includegraphics[width=\columnwidth]{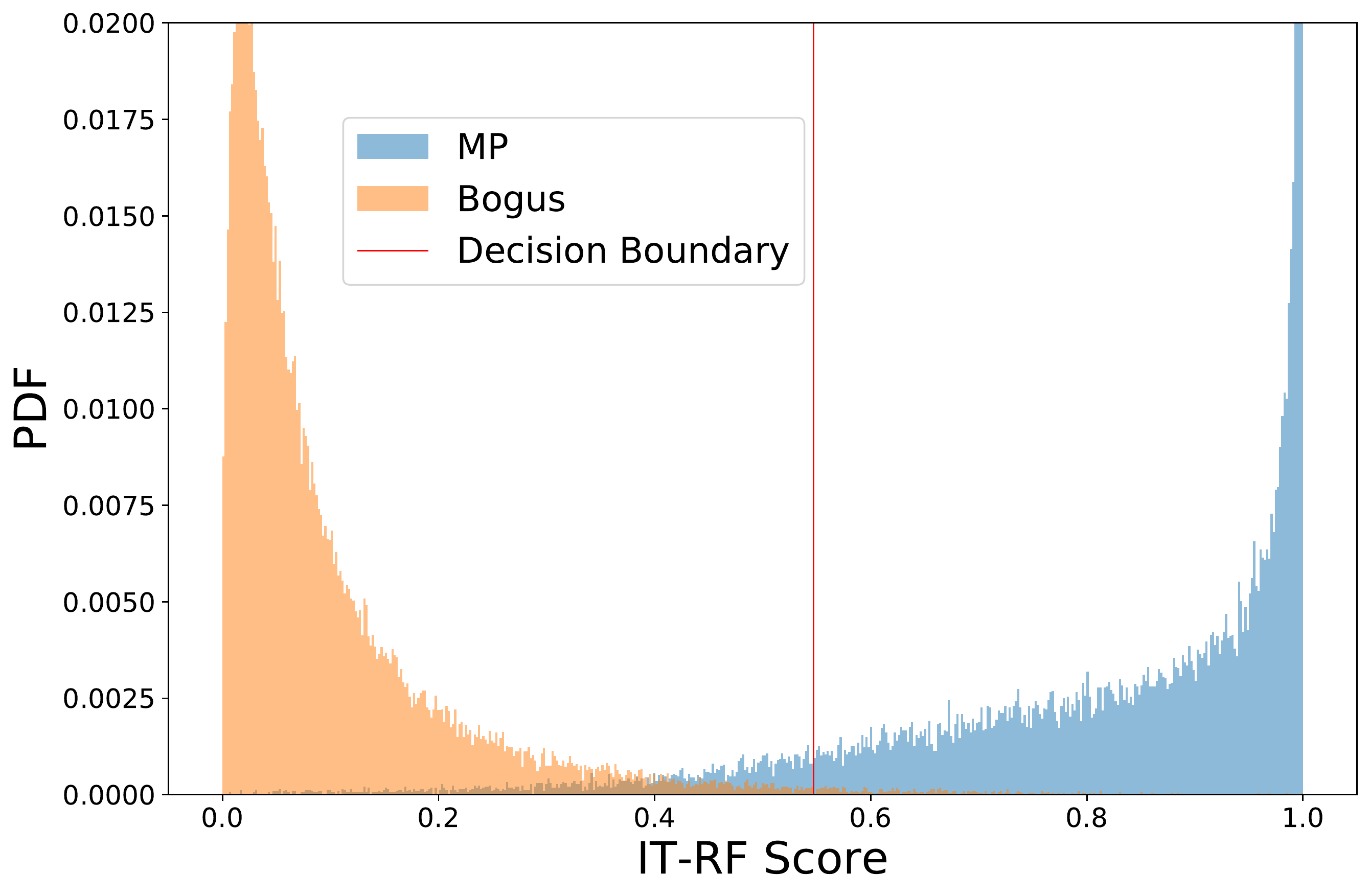}
    \caption{The classification score distributions of the MP test set. The \textit{top} and the \textit{bottom} plots represent the QB-RF and the IT-RF classifiers respectively. The \textit{orange} histograms represent the score distribution of the bogus detections, meanwhile the \textit{blue} histograms represent the distribution of the MPs. The \textit{red} lines indicate the decision boundaries set at FPR=0.01.}
    \label{fig:rb_distribution}
\end{figure}

Fig. \ref{fig:rb_distribution} shows that both QB-RF and IT-RF classifiers can separate bogus and real detections in the MP test set effectively. There are small overlapping regions at around 0.5 for the QB-RF distribution and 0.4 for the IT-RF distribution. The difference between these two decision boundaries is caused by the ratio difference between the numbers of the real and the bogus samples in the training sets.
\begin{figure}
	\includegraphics[width=\columnwidth]{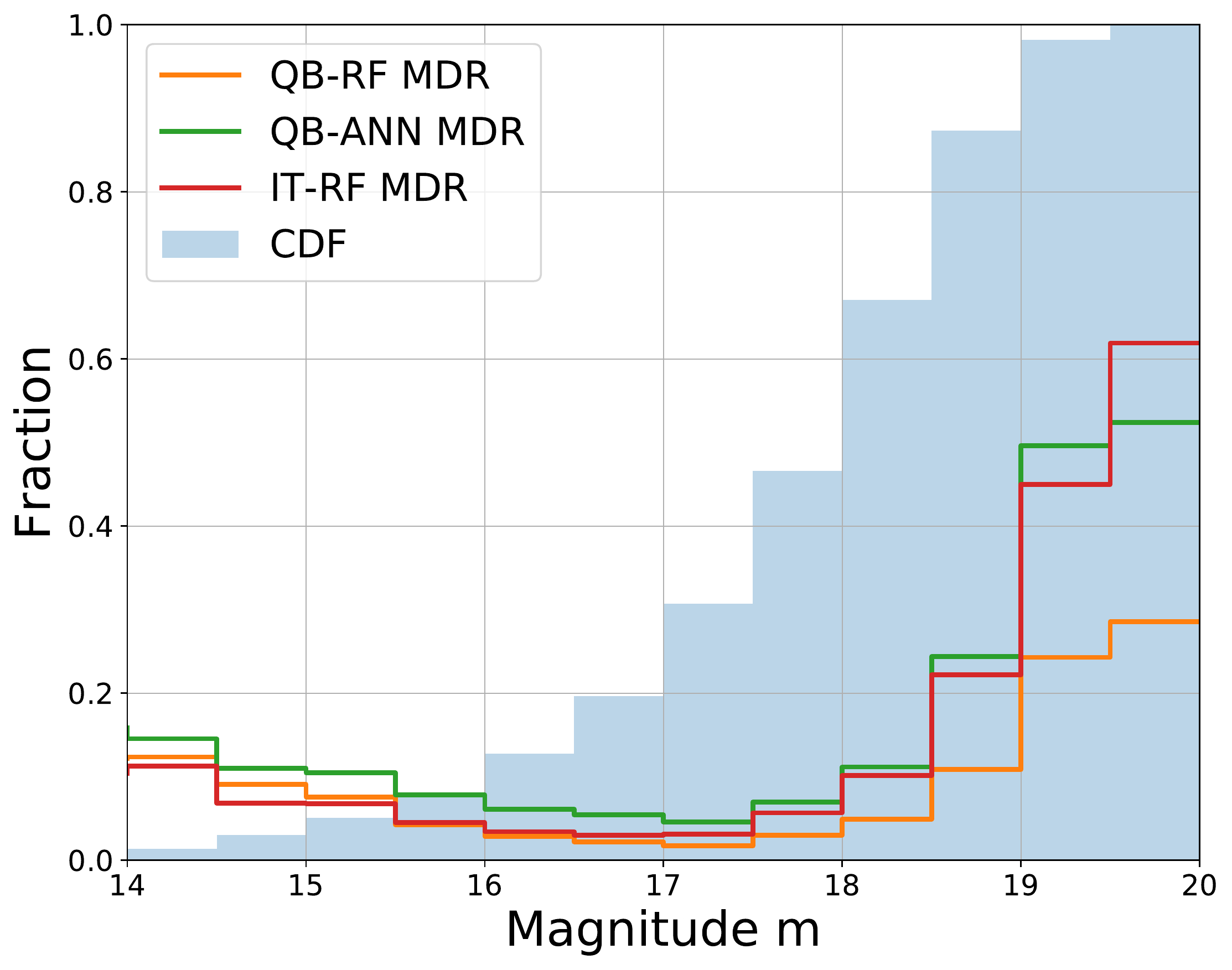}
    \caption{MDR for the MPs as a function of magnitudes. The QB-RF classifier indicated by the \textit{orange} line always shows the lowest MDR. The MDRs of QB-ANN and IT-RF are also plotted with \textit{green} and \textit{red} lines, respectively. The \textit{blue} blocks show the cumulative distribution function (CDF) with magnitude of the MPs.}
    \label{fig:mp_mdr_mag}
\end{figure}

We can see that the results showed in Figs. \ref{fig:mp_roc} and \ref{fig:mp_mdr_mag} are different from what the Figs. \ref{fig:roc} and \ref{fig:mdr_mag} present. The QB-RF shows a much lower FoM of about 5\% in Fig. \ref{fig:mp_roc} than in Fig. \ref{fig:roc}. However, the conclusions which can be drawn from both ROC curves are the same, the QB-RF performs the best in terms of both FoM and AUC. The MDR-mag plots, Fig. \ref{fig:mp_mdr_mag} shows that the MDR of the QB-RF always stays below 0.3 even up to $m$=20, which is much lower than the one of $>0.9$ in Fig. \ref{fig:mdr_mag}. 
There are several potential factors causing these differences. First, we can see the CDFs in Figs. \ref{fig:mdr_mag} and \ref{fig:mp_mdr_mag} are different, indicating two different brightness distributions of the real samples in the data sets. In the injection data set, we inject sources with a uniform brightness distribution. On the other hand, in the MP test set, there is a more accurate representation of the generalized magnitude distribution in comparison to the artificial one from our injection set. Second, we only use the images taken in a particular field with particular instruments, UT3 and UT4, to build our injection set. In contract, the MP test set includes detections from images taken with a wider range of conditions, with different UTs, fields, image quality scores, etc. Finally, the PSF models used to generate the injections can never be fully representative of the range of genuine detections appearing on the difference images as they are discretized on the image.

We provide evidence that the training set constructed using our quick-building strategy is not only fast and convenient, but shows nearly identical performance to the classifiers trained in the traditional way. Since the main scope of this paper is to show how to address the problem of assembling a sufficiently large data set for supervised machine learning, the performance comparison between the different learning algorithms is for reference only. The results might depend on the architectures of the classifiers, the feature representation, etc.

\subsection{Feature importance}
\begin{figure}
    \centering
	\includegraphics[scale=0.5]{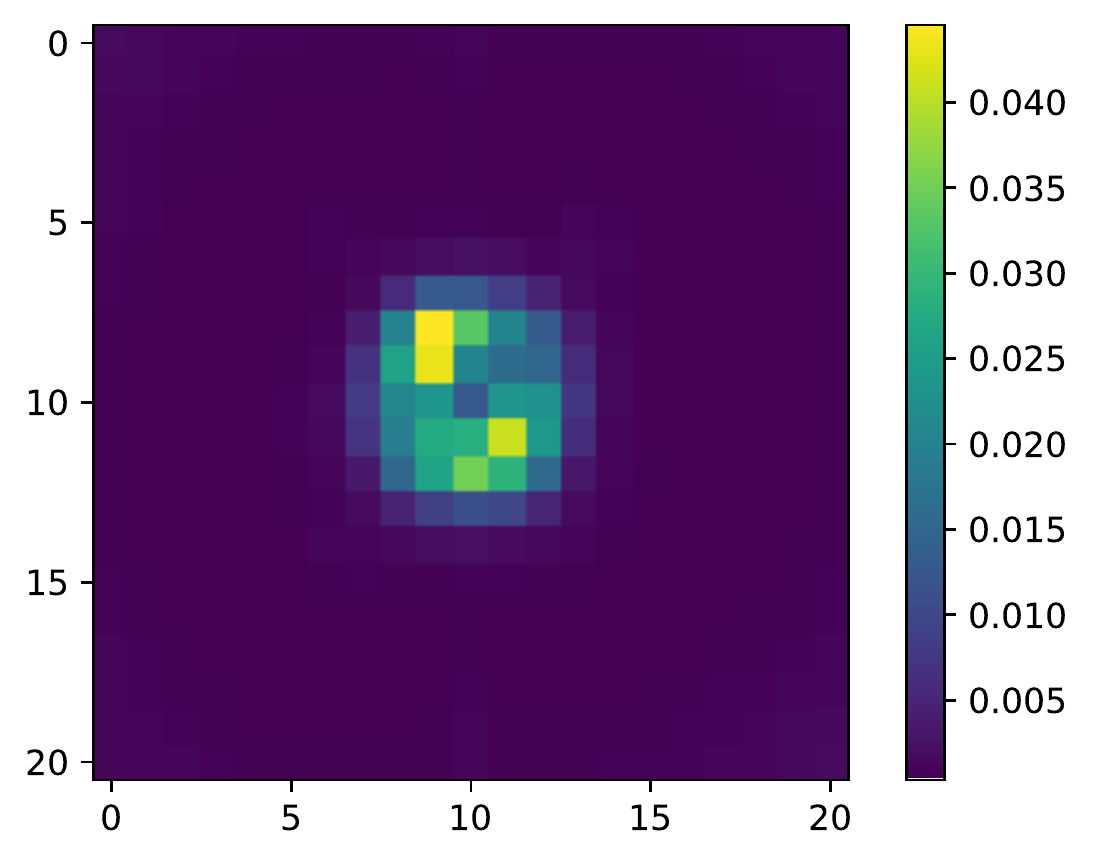}
    \caption{The RF feature importance of each pixel over the stamp. It shows that the central 7-by-7 pixels are the most important features for separating real and bogus detections using our QB-RF classifier.}
    \label{fig:feature_imp}
\end{figure}

To understand how the RF classifier calculates the classification score for a detection thumbnail, we can simply plot out the feature importance of each pixel (see Fig. \ref{fig:feature_imp}). As we expect, to classify whether a detection is real or bogus, the classifier only considers the central 7-by-7 pixels as the most important features. This area is consistent with the 90\%-percentile of the FWHM distribution, which is 7.8 pixels for GOTO prototype performance, for the real samples in our QB training set.

Fig. \ref{fig:feature_imp} shows that the central pixel is not the most important pixel feature among the entire stamp. This could be due to the elongation of the PSF of the real detections we used to train our classifier (see Fig. \ref{fig:example_stamps}).

Additionally, the pixels outside the 7-by-7 central region have very low values of feature importance. There are two conclusions that can be drawn from this observation. The classification scores for those transients close to bright objects or galaxies would not be affected. However, the subtraction residuals from the bright objects could easily be scored with a high value. Fortunately, the subtraction residuals due to the bright objects can easily be filtered by human vetting which should always be done as a confirmation of the candidates after the automatic real-bogus classification process. Another method of solving this problem is to reject candidates within a certain angular distance from  bright objects.
\begin{figure}
    \centering
	\includegraphics[width=\columnwidth]{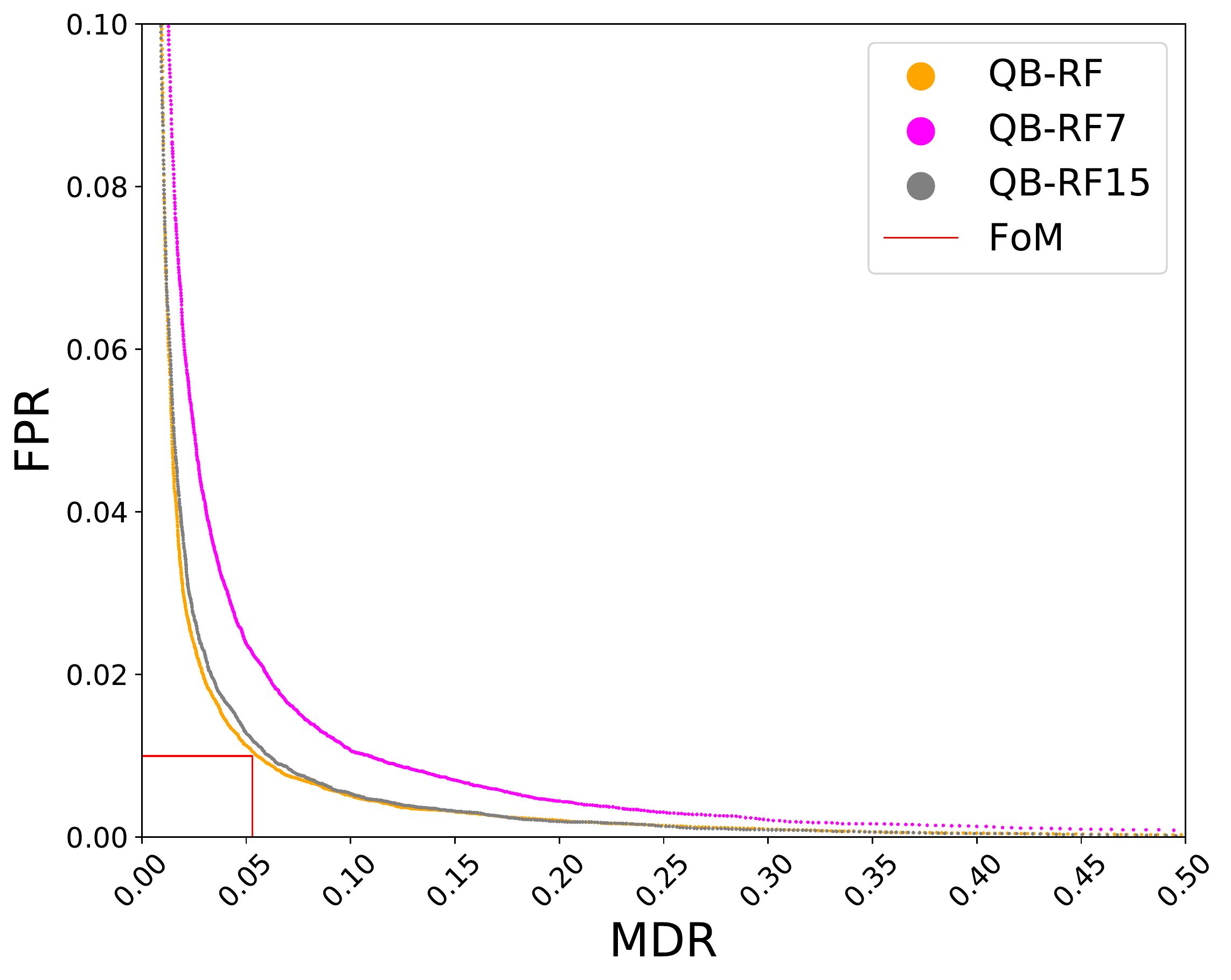}
    \caption{The ROC curves of the QB-RF ({\it orange}), the QB-RF7 ({\it magenta}) and the QB-RF15 ({\it grey}) classifiers tested on the MP test set. The QB-RF7 doubles the FoM compared to the QB-RF. Both QB-RF and QB-RF15 perform consistently.}
    \label{fig:7x7_compare}
\end{figure}

The feature importance of our RF model prompted us to train another classifier with a different stamp size. We used stamp sizes of 7-by-7 and 15-by-15 pixels to train additional models (called QB-RF7 and QB-RF15). Since we use the median pixel value on each stamp as the noise level to perform scaling and filling the masked pixels, if the stamp size is close to the PSF area, the median pixel value may not well represent the noise level. Therefore, we use the original 21-by-21 pixel stamp to obtain the noise level, and then use another crop to generate a smaller pixel stamp for our training features.

We use the MP test set only to compare the differences between models trained with different stamp sizes. Figure \ref{fig:7x7_compare} shows that the FoM of the QB-RF7 is about $10\%$, which is about twice that for QB-RF, but the ROC curve of the QB-RF15 is consistent with the QB-RF classifier. Therefore, we suggest using stamp sizes of at least twice the $90\%$-percentile of the FWHM for training.

\section{Conclusion and summary}\label{sec:conclusion}
In this paper, we design and test methods to separate real detections in optical difference imaging from bogus ones,  by using machine learning methods. Manually building a large training set is very time consuming, which motivates the use of detections in the science images, which should look identical to transients in the subtracted images, as the real sample. Our training set consists of 400\,000 real and bogus detections, respectively. We use scaled pixel values over a 21-by-21-pixel stamp centered at the detection position to represent the features of each detection to calculate the real-bogus score.

The RF classifier is shown to have a better performance compared to ANN by testing with the MP data set. We obtain an overall accuracy of 97.1\% and FoM of 5.2\% with the decision boundary set to 0.61. We also show that the classifier trained on our quick-build training set has a similar performance with the classifier trained on our injection data set.

Compared to the traditional methods used to build a training set for supervised machine learning methods, our strategy can help to build a training set of reasonable size within few days without having to spend weeks to months on manual inspection and human verification. We also show that the performance of the classifier built based on this strategy is comparable to the classifier built by traditional methods. 

We also build two other RF classifiers by training on 7-by-7 and 15-by-15 pixel stamps, to study how the performance varies with stamp size. We show that a 15-by-15 pixel stamp is sufficient to train our model. Therefore, we recommend using at least twice the $90\%$-percentile FWHM as the training stamp size.

While the quick-build strategy we use to build our training set is both fast and effective to train our classifier, we do not prescribe this technique to assess the best method of building a classifier overall. Instead, we suggest it could serve as a preliminary classifier for transient searches with newly-operational optical telescopes, or being ideal for small research collaborations that decide to pursue transient search projects. Since we only use the pixel intensity for performing classification, the idea of this work, in principle, should be directly applicable with other instruments.

\section*{Acknowledgements}
We thank the referee for the comments that improved this
paper. The Gravitational-wave Optical Transient Observer (GOTO) project acknowledges the support of the Monash-Warwick Alliance; Warwick University; Monash University; Sheffield University; University of Leicester; Armagh Observatory \& Planetarium; the National Astronomical Research Institute of Thailand (NARIT); the Instituto de Astrof\'isica de Canarias (IAC) and the University of Turku. RB, MK and DMS acknowledge support from the ERC under the European Union's Horizon 2020 research and innovation programme (grant agreement No. 715051; Spiders).

\section*{Data availability}
{Data products will be available as part of planned GOTO public data releases.}



\bibliographystyle{mnras}
\bibliography{all.bib} 








\bsp	
\label{lastpage}
\end{document}